\documentclass[prb,twocolumn,showpacs,superscriptaddress,floatfix]{revtex4}
\usepackage{graphicx}

\begin{document}
\title{Co-operative density wave and giant spin gap in
the quarter-filled zigzag ladder}
\author{R.T. Clay}
\affiliation{Department of Physics and Astronomy and ERC Center for Computational 
             Sciences, Mississippi State University, Mississippi State MS 39762}
\author{S. Mazumdar}
\affiliation{ Department of Physics, University of Arizona
Tucson, AZ 85721}
\date{\today}
\begin{abstract}
Strong co-operative interactions occur between four different broken
symmetries involving charge-ordering and bond distortions in the
quarter-filled correlated zigzag electron ladder. The ground state is
singlet, with spin gap several times larger than in the spin-Peierls
state of the one-dimensional quarter-filled chain with the same
parameters. We propose the quarter-filled zigzag electron ladder model
for several different organic charge-transfer solids with coupled
pairs of quasi-one-dimensional stacks, the spin-gap transition
temperatures in which are unusually high.
\end{abstract}
 \maketitle

A characteristic feature of one-dimensional (1D) metals is the Peierls
transition, where electron-phonon (e-ph) interactions lead to the
opening of energy gaps in charge and spin degrees of freedom (DOF) and
an insulating ground state.  In the presence of strong
electron-electron (e-e) repulsive interaction, electronic Hamiltonians
describing a 1/2-filled band may be reduced to an antiferromagnetic
Heisenberg spin Hamiltonian in which the charge DOF are absent.  With
nonzero spin-phonon coupling a spin gap (SG) appears again in the
so-called spin-Peierls (SP) state, in which there occurs lattice
dimerization accompanied by the formation of singlet spin-bonded pairs
of electrons. This mechanism of SG formation is absent in the two
dimensional (2D) 1/2-filled band, where antiferromagnetism rather than
the SP state dominates.

Away from 1/2-filling, charge DOF are important over and above the
spin DOF, and the SP transition in 1D occurs only after a
metal-insulator (M-I) transition freezes the charge DOF within an
electronic Hamiltonian. The most studied case is that of the
1/4-filled chain, where the SP transition is accompanied by charge
ordering (CO) \cite{Clay}.  Consider the 1D extended Peierls-Hubbard
model,
\begin{eqnarray}
H_{1D}&=& -\sum_{j} [t-\alpha\Delta_{j,j+1}]B_{j,j+1} + 
\frac{K}{2}\sum_{j}{(\Delta_{j,j+1})^2}
\nonumber \\
&+& U\sum_{j} n_{j,\uparrow}n_{j,\downarrow} + V\sum_{j} n_jn_{j+1}. 
\label{eqn-1d}
\end{eqnarray}
In the above $B_{ij}$ = $\sum_{\sigma}
(c^\dagger_{i,\sigma}c_{j,\sigma} + c^\dagger_{j\sigma} c_{i,\sigma})$
with $\sigma$ the electron spin, $\alpha$ and $K$ are the intersite
e-ph coupling and spring constant, respectively, $\Delta_{j,j+1}$ is
the Peierls distortion of the $j$th bond, and $n_j =
\sum_{\sigma}c^\dagger_{j,\sigma}c_{j,\sigma}$ is the total number of
electrons on site $j$. $U$ and $V$ are on-site and nearest neighbor
(n.n) Coulomb interactions.  Within this model, for $V < V_c(U)$ the
SP bond modulation is accompanied by 2k$_F$ CO, with site charge
densities going as $0.5+\epsilon$ , $0.5+\epsilon$, $0.5-\epsilon$ ,
$0.5-\epsilon$ . We shall refer to this CO pattern as ...1100..., or
as a bond-charge density wave (BCDW) \cite{Clay}.  For $V > V_c(U)$,
the M-I transition is to the CO state ...1010..., and the SP
transition involves dimerization of the bond distances between the
``occupied'' sites.  These two possible CO
\begin{figure}[tb]
\centerline{\resizebox{3.4in}{!}{\includegraphics{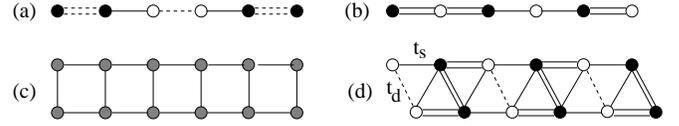}}}
\caption{(a) The 1D ...1100... BCDW SP state. Black (white) circles
 represent large (small) charges. Bond strengths decrease as solid
 bond $>$ double dotted $>$ single dotted.  (b) 1D 1/4-filled SP state
 with ...1010.. CO.  (c) The 1/4-filled rectangular ladder.  (d) BCDW
 state in the 1/4-filled zigzag ladder. Bonds are alternately strong
 and weak along the stacks, and have periodicity 4 along the zigzag
 interstack direction. }
\label{BCDW}
\end{figure}
states are shown in Figs.~1(a) and (b).  The SP transition
temperatures in 1/4-filled band quasi-1D organic charge-transfer
solids (CTS) are typically in the range 10 -- 20 K
\cite{Pouget97,Visser83}.

Intermediate between 1D and 2D are ladder lattices, consisting of two
or more coupled 1D chains.  We will distinguish between the two most
common ladder geometries, {\it rectangular} (see Fig.~1(c)) and {\it
zigzag} (see Fig.~1(d)).  The zigzag ladder is sometimes considered
also as a 1D chain with n.n as well as next-nearest neighbor (n.n.n.)
couplings. The properties of ladder systems have been pursued in great
detail in the 1/2-filled electron band or spin-model case. SG are
found in antiferromagnetic S = 1/2 ladders for both even-leg
rectangular\cite{Dagotto96} and zigzag systems\cite{Majumdar69}. SGs
and superconducting pair-pair correlation functions in weakly doped
even-leg rectangular ladders have been investigated within t-J and
Hubbard models \cite{Dagotto99}.  Away from 1/2-filling, charge and
spin orderings in the 1/4-filled rectangular electron ladder have been
investigated within an extended Hubbard Hamiltonian with nonzero
n.n. Coulomb interaction \cite{Vojta}.  There a zigzag CO pattern, in
which single electrons occupy opposite vertices of n.n. rungs, occurs
for sufficiently large n.n Coulomb interaction $V>V_c$ \cite{Vojta}.
SG is possible in this CO phase, but with very small
magnitude\cite{Vojta}.

In the present Letter, we examine the ground state broken symmetries
in the 1/4-filled band zigzag electron ladder.  We show that because
of the strong cooperative interaction between multiple broken
symmetries, SGs considerably {\it larger} than those found in the
1/4-filled single chain are possible.  We then point out that that
this state is likely realized in certain organic CTS with coupled
chains.

To understand the BCDW and cooperative effects in the 1/4-filled
zigzag ladder, we first consider the appropriate one-electron
Hamiltonian which includes intersite e-ph interactions.
\begin{eqnarray}
H^{1e}_{zz}&=&  -\sum_{{\langle ij \rangle},\sigma} [t_s-\alpha_s\Delta^s_{ij}]B^s_{i,j} + \frac{K_s}{2}\sum_{i}{(\Delta^s_{ij})}^2 \nonumber \\
&-&\sum_{{[ij]},\sigma} [t_d-\alpha_d\Delta^d_{ij}]B^d_{i,j} +
\frac{K_d}{2}\sum_{i}{(\Delta^d_{ij})}^2 
\label{eqn-1ezz}
\end{eqnarray}
In the above, subscripts $s$ and $d$ label respectively intrastack and
diagonal zigzag interstack hopping integrals and bond distortions (see
Fig.~1(d)), and $\langle \cdot\cdot \rangle$ and $[\cdot\cdot]$ denote
intrastack and interstack n.n. bonds. The broken symmetry of Fig.~1(d)
is then simply the 2k$_F$ Peierls distortion that is expected when the
zigzag ladder is considered as a 1D chain with n.n. hopping $t_d$ and
n.n.n. hopping $t_s$.  Unconditional Peierls transition occurs for
$t_d >$ 0.5858 $t_s$. We have chosen $t_d/t_s$ = 0.7.  Such large
$t_d$ have been estimated for the experimental CTS we are interested
in and are due to the close interchain contacts involving sulfur atoms
\cite{Ribera99,Rovira00,Nakamura02}. An interesting aspect of this
Peierls distorted state is that the expected period 4 2k$_F$ Peierls
bond and charge distortions along the zigzag interstack direction are
accompanied by period 2 distortions of the same quantities along the
stacks. This has important ramifications in the presence of e-e
interactions, as we show below.

We now ask whether this broken symmetry state persists when Coulomb
interactions are added.
To the one-electron Hamiltonian of Eq.~2 we add an additional term
containing e-e interactions:
\begin{equation}
\label{eqn-phham}
H_{zz}^{ee} = U\sum_{i} n_{i,\uparrow}n_{i,\downarrow} + 
V_s\sum_{\langle ij \rangle} n_in_j + V_d\sum_{[ij]} n_in_j
\label{eqn-zz}
\end{equation}
$U$, $V_s$, $V_d$ are the on-site, intrastack n.n., and interstack
n.n. Coulomb interactions. 
\begin{figure}[tb]
\centerline{\resizebox{3.5in}{!}{\includegraphics{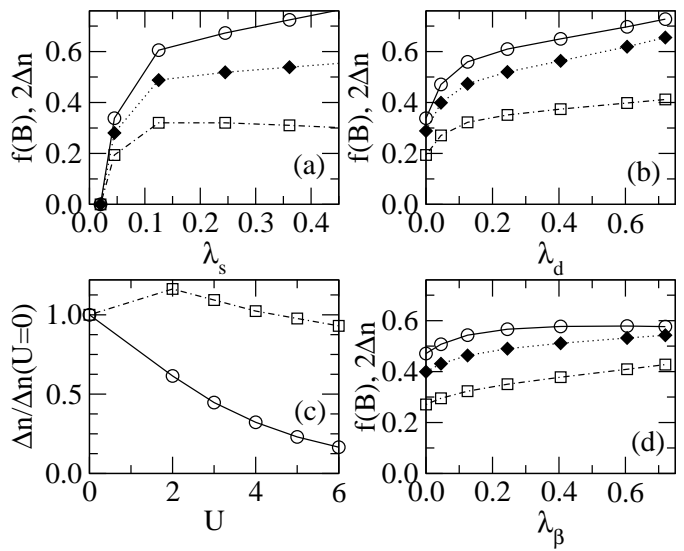}}}
\caption{(a) and (b). Self-consistent order parameters for the
1/4-filled 16-site periodic zigzag electron ladder (see text).  Open
circles, filled diamonds and open squares correspond to $f(B_s)$,
$f(B_d)$ and 2$\Delta n$.  (c) Normalized $\Delta n$ versus $U$ in the
16-site 1D ring with n.n. only hopping (circles), and for the zigzag
cluster (squares).  $V_s$ = $V_d$ = 0 for $U = 0$, $V_s$ = 1.0 for all
nonzero $U$ in both the 1D ring and the zigzag cluster, $V_d$ = 0.5 in
the zigzag cluster. (d) Order parameters $f(B_s)$, $f(B_d)$ and
2$\Delta n$ versus Holstein coupling $\lambda_\beta$.}
\label{exact}
\end{figure}
We have investigated Hamiltonian $H=H_{zz}^{1e}+H_{zz}^{ee}$
numerically,
performing exact diagonalization calculations for 16-site periodic
zigzag clusters and Constrained Path Quantum Monte Carlo (CPMC)
\cite{CPMC} calculations of long (up to 128 sites) open as well as
periodic zigzag ladders. Investigating the complete phase space would
require varying three Coulomb interaction parameters, two hopping
integrals as well as the different e-ph couplings. We have therefore
restricted ourselves to the parameter regime appropriate for organic
CTS \cite{Clay}, {\it viz.}, $U > 4t_s$, $t_s > t_d$, $V_d < V_s <
U/2$.  In what follows, we have taken $t_s$ = 1.  The procedure for
the self-consistent exact diagonalization is the same as in
Refs.[\onlinecite{Clay,Riera00}] for 1D periodic rings with
n.n. hopping. For the range of parameters above the ground state we
find is either completely uniform, or is the broken symmetry state of
Fig.~1(d).  We evaluate self-consistently all $\langle B^s_{ij}
\rangle$, $\langle B^d_{ij} \rangle$ and $\langle n_i \rangle$ and
define order parameters (i) $f(B_s)$, the absolute value of the
difference between consecutive $\langle B^s_{ij} \rangle$ along the
stacks, divided by the average $\langle B^s_{ij} \rangle$, (ii)
$f(B_d)$, the difference between the strongest and the weakest
$\langle B^d_{ij} \rangle$ along the zigzag direction, divided by the
average $\langle B^d_{ij} \rangle$, and (iii) $\Delta n$, the
difference between the charge densities of consecutive sites along the
stacks.  {\it Co-operative broken symmetry requires that all three
order parameters are simultaneously nonzero}.

 Fig.~2(a) shows the behavior of the three order parameters for fixed
$U = 6, V_s = 1, V_d = 0.5$, as a function of the dimensionless e-ph
coupling $\lambda_s = \alpha_s^2/K_st_s$, with the other e-ph coupling
$\lambda_d = \alpha_d^2/K_dt_s$ held at 0.  All three order parameters
simultaneously become nonzero at $\lambda_s=\lambda_s^c = 0.045$.
$\lambda_s^c$ should be $0^+$ in the thermodynamic limit for the
transition to be unconditional, suggesting that it should decrease
with increasing size of finite clusters; we verified that
$\lambda_s^c$ is indeed smaller in 16 site compared to the 8 site
zigzag clusters (for 8 sites, $\lambda_s^c = 0.245$).  Similar
behavior are also seen in Fig.~2(b) as a function now of $\lambda_d$,
for fixed $\lambda_s = 0.045$.  Taken together, Figs. 2(a) and (b)
clearly indicate the co-operative nature of the transition to the BCDW
state in the 1/4-filled zigzag ladder.

The broken symmetries within the correlated zigzag ladder are far
stronger than in 1D. To demonstrate this we compare $\Delta n$ in the
two cases as a function of $U$ with e-ph coupling $\lambda =
\alpha^2/Kt$ = 1.28 for 1D and $\lambda_s$ = 0.125, 
$\lambda_d$ = 0 for the zigzag ladder. The distortion amplitudes
decrease rapidly with $U$ in 1D (see below), which necessitates the
larger $\lambda$ in this case.  In Fig.~2(c) we have plotted the
$\Delta$n normalized by its value for $U$ = 0 for both cases. For all
nonzero $U$ we have chosen $V$ = 1 for the 1D ring and $V_s$ = 1,
$V_d$ = 0.5 for the zigzag cluster.  The distortion in the zigzag
ladder is affected very weakly by e-e interactions because of the
co-operative nature of the transition here.  As pointed out above, the
charge and bond periodicities along the stack direction in Fig.~1(d)
are 4k$_F$, even as these have 2k$_F$ periodicities along the zigzag
interstack direction. In purely 1D the 4k$_F$ distortions get stronger
with e-e interactions, while the 2k$_F$ distortions get weaker. There
is thus a tendency to cancellation of these effects in the zigzag
cluster, with the distortion amplitude remaining the same.

As one goal of our work is to apply our theory to organic CTS with
molecular sites we have also investigated 
the effect of Holstein electron-molecular vibration coupling:
\begin{equation}
H_{\beta} = \beta\sum_{i}v_{i}n_{i} + \frac{K_{\beta}}{2} \sum_{i}v_{i}^2 
\end{equation}
Here $\beta$ is the intrasite e-ph coupling with corresponding spring
constant $K_{\beta}$, and $v_i$ is the amplitude of the internal
molecular vibration.  In Fig.~2(d) we have plotted the order
parameters for $H=H_{zz}^{1e}+H_{zz}^{ee}+H_{\beta}$ against
$\lambda_{\beta} = \beta^2/K_{\beta}t_s$ , with fixed $\lambda_s$ =
$\lambda_d$ = 0.045. All three order parameters again increase with
$\lambda_{\beta}$.  Our conclusions regarding cooperative broken
symmetries remain the same whether or not the Holstein interaction is
included.

We have verified the coexisting broken symmetries for larger lattices
by performing CPMC calculations for long zigzag clusters with open
boundary condition (OBC).  In 1D, the central region of long {\it
open} chains exhibit spontaneous charge or bond-order distortions,
{\it even with uniform hopping integrals and zero e-ph couplings}
\cite{Clay01}.  We have performed CPMC calculations for a 64-site
zigzag open ladder with $\lambda_s = \lambda_d = \lambda_{\beta} = 0$,
and for the same $U, V_s$, $t_s$ and $t_d$ as in Fig. 2(a). With OBC,
$f(B_s), f(B_d)$ and $\Delta n$ depend on the locations of the sites
being considered, and we have therefore plotted the charge densities
$\langle n_i \rangle$ and the bond orders $\langle B^s_{ij} \rangle $
at chain centers in Fig.~3. Simultaneous 4k$_F$ dimerizations of
charge and bond along the stacks (Figs. 3(a) and (b)) and simultaneous
2k$_F$ modulations of these quantities along the diagonal zigzag
direction (Figs.~3(c) and (d)) are seen, in complete agreement with
the results of Fig.~2.
\begin{figure}[tb]
\centerline{\resizebox{3.4in}{!}{\includegraphics{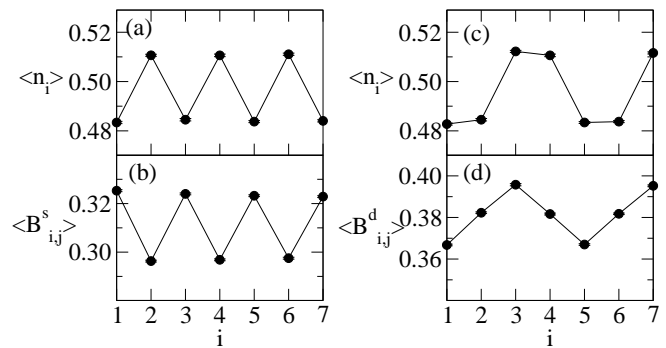}}}
\caption{Charge densities and bond orders in an open 64-site
1/4-filled zigzag ladder, as computed by CPMC. Site indices $i$ are
arbitrary and correspond to the central region of the open ladder,
along the stacks in (a) and (b), and along the interstack zigzag bonds
in (c) and (d).}
\label{open}
\end{figure}

The SG in the distorted zigzag 1/4-filled ladder is due to the      
formation of a strong 1-1 local singlet bond in the ...1100... zigzag BCDW.
We have calculated the SG in the BCDW state of the zigzag ladder using
CPMC. Our calculations are for long {\it periodic} zigzag clusters,
for the same parameters as in Fig.~3.  For comparison, we also
evaluate the SGs in the ...1100... BCDW state of a 1D periodic ring
(Eq.~(1), Fig.~1(a)), with $U$ = 6, $V$ = 1, $t$ = 1, using the
Stochastic Series Expansion Monte Carlo method \cite{Syljuasen02a}.
The reason for choosing the ...1100... state over the
...1010... CDW-SP state is that the SG in the former, with shorter 1-1
bond, is larger.  Finite periodic clusters do not have spontaneous
distortions while self-consistent CPMC calculations are not possible
for the large clusters we need to investigate to obtain the SGs in the
thermodynamic limit. Thus the BCDWs can be generated only by
externally imposing bond or charge distortions. The co-operative
nature of the zigzag BCDW demonstrated in Figs. 2(a) imply that either
of the two bond distortions or the CO is suitable for our purpose, as
the other two broken symmetries will be generated spontaneously.
Since the modulations of the hopping integrals are different in the 1D
chain and the zigzag ladder, we choose to externally impose COs that
are congruent with Figs.~1(a) and (d). This is achieved by adding a
site energy component $\sum_i \epsilon_i n_i$ to the $\lambda_s =
\lambda_d = 0$ limit of the Hamiltonian in Eq
(\ref{eqn-zz}). The site energies added are $+|\epsilon|, +|\epsilon|,
-|\epsilon|, -|\epsilon|$ along the zigzag bonds as in Fig.~1(d) and
along the linear chain as in Fig.~1(a).  We evaluate the
singlet-triplet gaps $\Delta_{\sigma}$ using CPMC for periodic zigzag
ladders with $N$ = 32, 64 and 128 sites, and for periodic 1D rings
with $N$ = 32, 64 and 96 sites, for several different $\epsilon$. For
each $\epsilon$, the $\Delta_{\sigma}$ and $\Delta n$ at $N \to
\infty$ are found from extrapolations against $1/N$.  In Fig.~4 we
have plotted the extrapolated $\Delta_{\sigma}$ against $\Delta n$ for
both the 1D system and the zigzag electron ladder.  The finite size
scaling of $\Delta_{\sigma}$ for the zigzag ladder for one value of
$\epsilon$ is shown in the inset.  For the same $\Delta n$,
$\Delta_{\sigma}$ is several times larger in the zigzag electron
ladder than in 1D ($\Delta_{\sigma}$ can be plotted also against
$|\epsilon|$; the same large difference between the zigzag ladder and
1D is obtained.)
\begin{figure}[tb]
\centerline{\resizebox{2.9in}{!}{\includegraphics{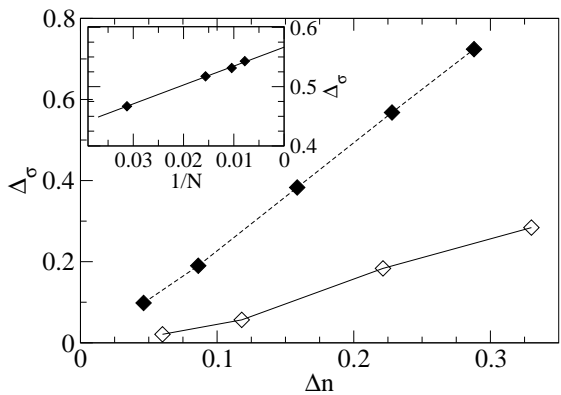}}}
\caption{Finite-size scaled spin gap $\Delta_{\sigma}$ versus $\Delta
n$ in the 1/4-filled band zigzag ladder (filled symbols) and in the 1D
chain (open symbols).  Lines are guides to the eye. Inset shows the
finite size scaling of $\Delta_{\sigma}$ in the zigzag ladder for
$|\epsilon|=0.3$. }
\label{SG}
\end{figure}
The large SG in the zigzag ladder is a direct consequence of the
strong local interstack 1-1 singlet bond, which in turn results 
from stronger distortions than in 1D.

Several recently discovered 1/4-filled band CTS are very likely
structurally zigzag ladders.  These systems consist of pairs of 1D
stacks of organic donor molecules \cite{holes}, with strong intrapair
interchain couplings, and weak interpair couplings
\cite{Rovira99a,Ise01}. They undergo M-I transition at T$_{M-I} >$ 200
K, followed by an insulator-insulator (I-I) transition which is
accompanied by the opening of a SG at lower temperature
T$_{SG}$\cite{Ribera99,Rovira00,Wesolowski03,Nakamura02}.  The
peculiarity of these paired-stack systems
\cite{Ribera99,Rovira00,Wesolowski03,Nakamura02} is that while T$_{SP}
\sim$ 10 -- 20 K in the 1D CTS \cite{Pouget97,Visser83}, T$_{SG}$
$\sim$ 70 K in (DT-TTF)$_2$Au(mnt)$_2$ \cite{Ribera99,Rovira00} and is
even higher at 170 K in (BDTFP)$_2$PF$_6$(PhCl)$_{0.5}$
\cite{Nakamura02}.  The very large T$_{SG}$, and therefore the large
SG in these systems are highly unusual.

To explain the large SG in these systems, an {\it effective} two-leg
rectangular spin-ladder model has been proposed
\cite{Ribera99,Rovira00,Wesolowski03,Nakamura02}.  It is known
experimentally that each individual 1/4-filled stack dimerizes below
T$_{MI}$ \cite{Ribera99}.  Within the dimerized rectangular ladder
lattice motif, if each dimer unit along the legs containing one
localized spin is assumed to be equivalent to an effective single site
(see Fig.~6 in Ref. \onlinecite{Rovira00}), then an 1/2-filled
rectangular spin ladder is obtained. For strong enough interstack spin
coupling, large SG is now obtained. The mapping from the 1/4-filled
dimerized rectangular electron ladder to the 1/2-filled spin ladder,
however, has not been formally proved.

We believe, however, that the zigzag electron ladder is the more
appropriate model for these systems for a number of reasons.  The
crystal structures of the materials (see Fig.~2(a) in
Ref. \onlinecite{Ribera99} and Fig.~2 in Ref. \onlinecite{Nakamura02})
indicate that each molecular site is coupled to two molecules on the
partner stack, as would occur in the zigzag ladder. Quantum chemical
calculations of hopping integrals\cite{Ise01} support this viewpoint.
Most importantly, it has been suggested from EPR linewidth studies
that transition to a CO state might be occurring in
(BDTFP)$_2$PF$_6$(PhCl)$_{0.5}$ \cite{Nakamura02}. This last result,
if correct, will be in agreement with the zigzag electron ladder
model.  The key difference between spin-ladders and 1/4-filled
electron-ladders is that CO is absent in the former, while it is a
prerequisite to large SG in the latter. Experiments have previously
demonstrated CO in 1D \cite{Nad00} as well as 2D \cite{Miyagawa00}
CTS, while very recently, the occurrence of the ...1100... BCDW has
been experimentally confirmed in a CTS \cite{Yamamoto05}.  We predict
that experimental investigation of the coupled-stack CTS will find
evidence for CO.

R.T.C. acknowledges support from Oak Ridge Associated Universities and
the ERC Center for Computational Sciences. Work at Arizona was
partially supported by NSF-DMR-0406604. We acknowledge useful
discussions with J.L. Musfeldt.


\begin{thebibliography}{99}
\bibitem{holes} The carriers in the donor stacks are holes and not
electrons. This does not alter the physics discussed here.
\end{thebibliography}
\end{document}